\documentclass[12pt,preprint]{aastex}   

\def\kpchi{\,h^{-1}{\rm {kpc}}}

\def\msunhi{\,h^{-1}{\rm M_\odot}}

\begin{document}

\title{Influence of baryonic physics on the merger time-scale of galaxies in N-body/hydrodynamical simulations}
\author{C. Y. Jiang$^{1,2}$, Y. P. Jing$^{1}$, W. P. Lin$^{1}$}
\affil{$^1$Key Laboratory for Research in Galaxies and Cosmology, Shanghai Astronomical Observatory,
Nandan Road 80, Shanghai, 200030, China}
\affil{$^2$
Graduate School of the Chinese Academy of Sciences, 19A, Yuquan Road, Beijing, China}

\begin{abstract}

Following our previous work(Jiang et al.(2008)), in which we studied
the merger time-scale of galaxies in a high-resolution cosmological
hydro/N-body simulation, we investigate the potential influence of
uncertainties in the numerical implementation of baryonic physics on
the merger time-scale. The simulation used in the previous work
suffers from the overcooling problem which causes the central galaxies
of large halos too massive. This may result in a shortened merger
time-scale compared to that in the real universe. We run a similar
simulation, but the stellar mass is significantly reduced to model
another extreme case of low stellar mass. Our result shows that the
merger time-scale is little affected by the star formation recipes,
except for the satellites in nearly radial orbits which show a 22
percent higher time-scale in the lower stellar mass case. Since the
radial orbits only account for a small part of the satellites' orbits,
the fitting formula in Jiang et al.(2008) is still
applicable to a reasonable accuracy, nearly immune to the uncertainty
in the baryonic physics.

\end{abstract}

\section{Introduction}

Structures form hierarchically in the LCDM universe with larger
objects being assembled through merging of smaller building
blocks. When a massive group is formed, its central galaxy takes up a
special position so that it can grow by accreting surrounding gas and
satellite galaxies.  These satellite galaxies gradually lose their
energy and angular momentum under the action of dynamical friction,
and finally sink to the center of the host halo (primary halo) to
merge with the central galaxy. To obtain the merger time-scale is
crucial to understanding the role of mergers in the growth of
galaxies. In theoretical computation of this time-scale, the formula
given by \cite{lacey93} is generally used. It is derived from
Chandrasekhar's formula for the dynamical
friction \citep{chandrasekhar43}.

In our previous work(\citet{jiang08}, hereafter J08), we used a
high-resolution N-body/hydro simulation to show that, this widely used
dynamical friction formula underestimates the time-scale of minor
mergers and overestimates that of major mergers. We then gave a new
fitting formula for the merger time-scale measured from the
simulation(refer to eq.(5) in J08),
\begin{equation}
  T_{\rm fit}=\frac{f(\epsilon)}{2C}\frac {m_{\rm pri}}{m_{\rm sat}}
  \frac {1}{\ln \Lambda}\frac{r_{\rm
vir}}{V_{\rm c}}\ .
\end{equation}
$C$ is a constant, approximately equal to 0.43, an $\epsilon$ is the
circularity parameter of the satellite's orbit. $f(\epsilon)$ is the
function that denotes the circularity dependence,
$f(\epsilon)=0.94\epsilon^{0.60}+0.60$, which is obtained by fitting
the two free parameters $a$ and $b$ in
$f(\epsilon)=a\epsilon^{b}+0.60$. ${V_{\rm c}}$ and $r_{\rm vir}$ are
the circular velocity and the virial radius of the primary halo
respectively, assuming it is an isothermal sphere. The Coulomb
logarithm is in the form of $\ln [1+(\frac {m_{\rm pri}}{m_{\rm
sat}})]$.  Therefore, the mass dependence is purely represented in the
mass ratio of the primary halo to the satellite halo $\frac {m_{\rm
pri}}{m_{\rm sat}}$. We showed that this fitting formula could
properly account for the mass dependence and the circularity
dependence.  \cite{lagos08} investigated the impact of this formula in
a semi-analytical model, finding that it caused a delay in the period
of maximum merger activity towards $z \sim 1.5$ compared to $z \sim
2.5$ obtained for \cite{lacey93}, produced an order of magnitude
increase in the disk instability frequency, and gave a slightly better
agreement with the observed morphological at the high-mass end
compared to the formulae given by \cite{lacey93} and \cite{boylan08}
who gave a different result from ours.  \cite{kang08} showed that this
improved galaxy merger time-scale was successful in reproducing the
V-band luminosity function of Milky-Way satellites.  \cite{petsch08}
found that using a mass-dependent Coulomb logarithm, similar to our
description, could model the dynamical friction of satellites in host
halos reasonably well.

However, complex baryonic physics relevant to the formation of
galaxies is still poorly understood, and can not be implemented
unambiguously in hydro simulations. Our simulation also suffers such
kind of problems, among which the overcooling problem is the most
probable one that affects our result. Gas overcooling in central
galaxies of massive halos exists in current hydrodynamical simulations
with star formation and supernovae feedback(e.g.\cite{borgani04};
\cite{saro06}).  This is due to the lack of AGN feedback to a certain
degree. A larger stellar mass either for the central or the satellite
galaxy may result in a shorter merger time-scale. For the central
galaxy, if it is more massive and thus more extended, the merger
remnant would probably be identified earlier. For the satellite
galaxy, its mass becomes crucial in determining the merger time-scale
when the dark matter is already severely stripped, and therefore a
more massive satellite would lead to earlier merger seen from
eq.(1). Will the potential uncertainties of the implemented baryonic
physics affect our results? In J08, we argued that these physical
processes could not alter the results from the theoretical point of
view. For one thing, dynamical friction becomes very efficient only
when satellites migrate to the central part of the primary halo, and
therefore satellites spend most of their time during the merging
process at the outer part of the primary halo except for some radial
orbits. Consequently, the merger time-scale is set by the conditions
at the outer part of the primary halo, and the central galaxy of the
primary halo cannot influence the results significantly. Furthermore,
tidal stripping is not efficient enough to strip a significant
fraction of satellites' dark matter at the outer part of the primary
halo.  Therefore, the stellar mass of satellites does not play a
decisive role either.  For another, semi-analytical results based on
N-body simulations (\cite{springel01}; \cite{kang05}) are in
qualitative agreement with ours, which indicates that our results are
not affected by the baryonic physics. In this paper, we will come to
this question again, quantifying the influence of the improperly
implemented baryonic physics with a hydro/N-body simulation.

\section{Method}

We run a cosmological hydro/N-body simulation and measure the merger
time-scale of galaxies from it.  The simulation and the way of
obtaining the merger time-scale are described in J08, and the readers
are referred to that paper for more details. Here we only give a brief
description. The simulation is run using the SPH code
Gadget-2(\cite{springel05}). The physical processes implemented in it
include radiative cooling, star formation in a subresolution
multiphase medium and galactic winds (\cite{springel03}). The same
cosmological parameters are adopted as those in J08, except that we
use a halved baryonic density, that is to say, $\Omega_{\rm b}=0.022$
compared to $\Omega_{\rm b}= 0.044$ in J08. Since the sum of dark
matter density and baryonic density remains the same, the large scale
structure and the halo mass function are not affected. Only the masses
of galaxies are reduced. Fig.1 shows the stellar mass function of
galaxies(solid line) compared to that in J08(dashed line) at z=0. The
stellar mass is generally reduced to 1/3 when $\Omega_{\rm b}$ is
halved. The nonlinearity of the star formation efficiency with
$\Omega_{\rm b}$ most probably results from the lowered gas density in
the halo center.

After dark matter halos are identified using friends-of-friends(FoF)
method, halo merger trees are then built by tracing the halos at z=0
back to z=2.0. Since the dynamical friction acts on satellite galaxies
that orbit around their central galaxies, only main branches of merger
trees are considered. To reduce the artificial effects caused by the
finite numerical resolution, J08 only keeps those satellite halos with
their central galaxies more massive than $2.0 \times 10^{10}
\msunhi$. Here we reduce this mass limit to $1.0 \times 10^{10}
\msunhi$, accounting for the halved baryonic mass density.

Galaxies are also identified with the friends-of-friends method, with
a linking length of $4.88\kpchi$. The merger time-scale of galaxies is
defined as the time elapsed between the moment when the satellite
galaxy first crosses the virial radius of the primary halo and the
moment when it finally coalescences with the central galaxy. The
merger is identified in such a way that, if the satellite galaxy and
the central galaxy begin to have the same descendant at one snapshot
and continue to have the same descendant in the following four
snapshots($\ge$0.5 of the dynamical time of a halo).  We use this
criterion to ensure that it is a real merger, not just a close flyby.

\section{Result}

As in J08, we find that the Coulomb logarithm $\ln(\Lambda)$ is better
represented by $\ln (1+m_{\rm pri}/m_{\rm sat})$ than by the other two
forms $\ln (m_{\rm pri}/m_{\rm sat})$ and $1/2\ln [1+(m_{\rm
pri}/m_{\rm sat})^2]$. Seen from eq.(1), the influence of the reduced
stellar mass on the merger time-scale can be represented in the
circularity function $f(\epsilon)$, which is obtained from the
measured merger time-scale in the simulation. If all mergers are used
in computing $f(\epsilon)$, there would be a selection bias against
those long-time mergers.  Due to the limited time that the simulation
stops at z=0, satellites with larger $f(\epsilon)$, hence longer
merger time-scale for the same $\epsilon$, haven't got enough time to
merge into their central galaxies till z=0. Therefore, the statistics
of $f(\epsilon)$ got in this way would be biased low. This problem is
particularly severe for larger $\epsilon$, as it takes a longer time
to merge on more circular orbits. To avoid this selection bias, we
need to construct a $complete$ sample in which all central-satellite
pairs are merged till z=0(the time our simulation stops).  That is to
say, galaxy pairs at higher redshift are preferred, but compromise
needs to be made to have enough statistics. As in J08, we constructed a
$complete$ sample of primary halos and satellites at the first 14
snapshots(redshift 1.55-2.0) with mass ratio greater than 0.1. Here
the completeness is $95.3\%$, slightly lower than that in J08.

Fig.2 shows the median value of $f(\epsilon)$(square points) together
with the best fitting curve (solid line). The original result in J08
is also shown with triangles and the dashed line. We see that the
discrepancy is only notable in the two lowest circularity bins, while
the difference is little in the most circular bin. Reducing the
baryonic content leads to a prolonged merger time-scale for galaxies
on relatively radial orbits, which makes the fitting curve
flatter. Satellites on relatively radial orbits spend more time in the
inner region of the halo where the tidal stripping is efficient, and
therefore central galaxies of these satellite halos play a more
important role in determining the merger times, with less massive
galaxies taking longer time to merge.  However, we find that it only
makes a difference of about $22\%$ at $\epsilon \sim 1$ where the
discrepancy is the largest. Note that our artificial cut in the baryon
budget causes galaxies to reduce their masses to $1/3$ of their
original value. In J08, the relevant mass range of the satellite
galaxies lies mostly around the characteristic mass of galaxies, where
the mass function conforms to the observed one. Therefore, the galaxy
masses are over-reduced by adopting a halved $\Omega_{\rm b}$. That is
to say, the $22\%$'s difference is likely overestimated for galaxies
with low circularities.  Furthermore, relatively radial orbits which
suffer the uncertain baryonic physics most only account for a small
part of all satellites' orbits. Consequently, our fitting formula as
shown in eq.(1) should be still valid to a reasonable accuracy.

\section{Conclusion}

In \cite{jiang08}, we studied galaxy mergers in a cosmological
hydro/N-body simulation, and presented a fitting formula for the
merger time-scale of galaxies. However, due to uncertainties in the
implemented baryonic physics, the simulation used in that paper
suffered from some shortcomings. The most probable problem that could
affect our result was the overcooling problem, which caused the
stellar mass of central galaxies in massive halos too high.  This may
produce a shortened merger time-scale compared to the real case in the
universe.  In this short paper, we have investigated the influence of
the uncertainty in the baryonic physics on the fitting formula. We
have modeled another extreme case of low stellar mass by artificially
reducing the baryon budget to its half in a cosmological hydro/N-body
simulation.  We find that, it makes a difference of $22\%$ at most for
low circularity orbits.  While for satellites in nearly circular
orbits, the merger time-scale is almost not influenced.  We note that
the stellar mass is reduced to one third of its original value in our
new simulation.  While it is an extreme case, the difference in the
stellar mass by three times can only produce a marginal difference in
the result. This indicates the robustness of our fitting formula to
different reasonable implementations of baryonic physics. Furthermore,
since low circularity orbits for which the discrepancy is the largest,
only account for a small part of the satellites' orbits, the fitting
formula in \cite{jiang08} is still applicable to a reasonable
accuracy, nearly immune to the uncertainty in the baryonic physics.

\acknowledgments
This work is supported by NSFC (10533030, 10821302, 10878001), by the
Knowledge Innovation Program of CAS (No. KJCX2-YW-T05), and by 973
Program (No. 2007CB815402).

\begin{figure}
\begin{center}
\plotone{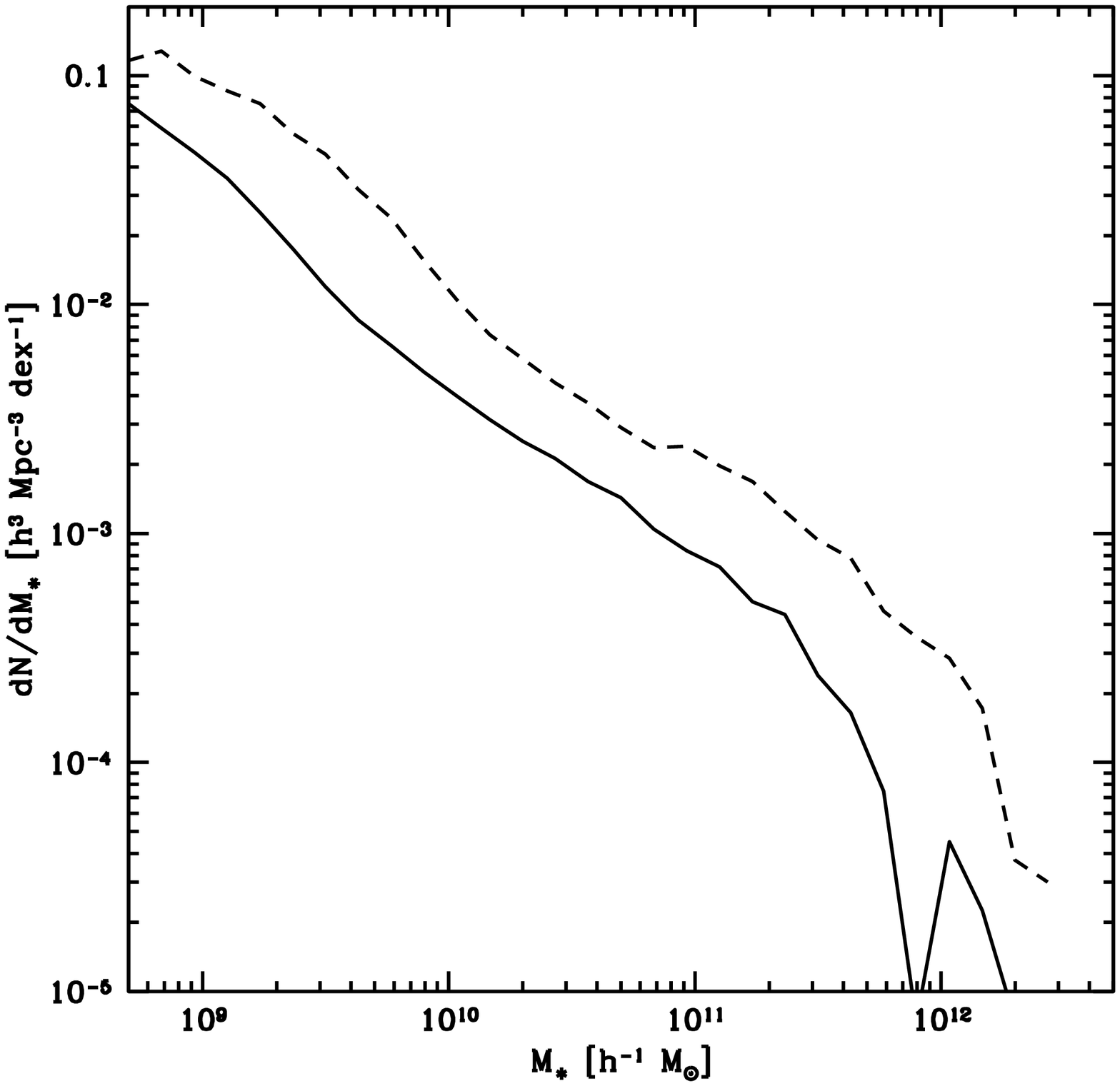}
\caption{Stellar mass function at z=0 in this work(solid line) and in J08(dashed line)}
\end{center}
\end{figure}

\begin{figure}
\begin{center}
\plotone{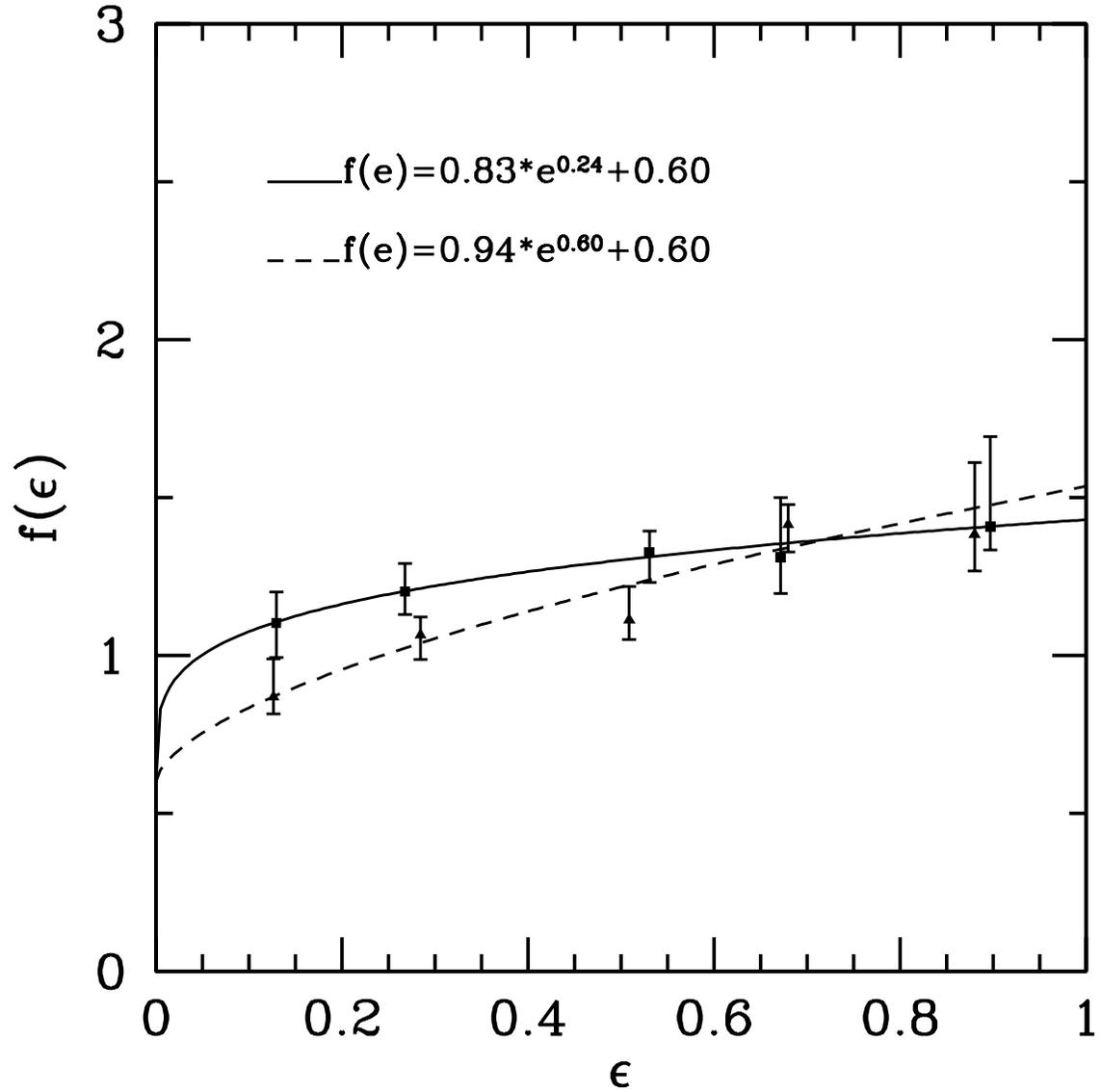}
\caption{Circularity function $f(\epsilon)$. Square points and the solid line show the result 
in the simulation where a halved $\Omega_{\rm b}$ is used. The dashed line represents the fitting
curve in J08.}
\end{center}
\end{figure}

\end{document}